\documentclass[runningheads]{llncs}
\usepackage{authblk}
\usepackage[square,numbers,sort&compress]{natbib}
\usepackage[a4paper, total={6.2in, 9in}]{geometry}
\usepackage[breaklinks=true,colorlinks=true,linkcolor=blue,urlcolor=blue,citecolor=blue]{hyperref}
\usepackage[justification=centering]{caption}
\usepackage{amsmath}

\usepackage{multicol, blindtext}
\usepackage{amssymb}
\usepackage[shortlabels]{enumitem}
\usepackage{qcircuit}
\usepackage{units}
\usepackage{tabu}
\usepackage[english]{babel}
\usepackage{algorithm}
\usepackage{dsfont}
\usepackage{color} 
\usepackage[framemethod=TikZ]{mdframed}
\usepackage{algorithm}
\usepackage{algpseudocode}
\usepackage{braket}
\usepackage{etoolbox}
\apptocmd{\sloppy}{\hbadness 10000\relax}{}{}
\usepackage{sectsty}
\usepackage{bold-extra}
\usepackage{physics}
\usepackage{tabu}

\usepackage{tikz}
\usetikzlibrary{positioning, shapes}

\pgfdeclarelayer{background}
\pgfdeclarelayer{foreground}
\pgfsetlayers{background,main,foreground}

\usepackage{graphicx}
\usepackage{tikz}
\usetikzlibrary{positioning}

\usepackage{soul}
\setstcolor{blue}

\usepackage[utf8]{inputenc}
\usepackage{pgfplots}
\pgfplotsset{compat=newest}
\usepgfplotslibrary{groupplots}
\usepgfplotslibrary{dateplot}

\usepackage{graphicx}
\usepackage{subcaption}
\usepackage{tabu}

\usepackage{tocloft}

\definecolor{ForestGreen}{RGB}{34, 139, 34}
\definecolor{DeepSkyBlue}{RGB}{0, 191,255}
\definecolor{Lavender}{RGB}{230, 230, 250}

\newcommand{\figref}[1]{Fig. \ref{#1}}
\renewcommand{\eqref}[1]{(Eq. \ref{#1})}

\newcommand{\comment}[1]{\footnote{}}

\DeclareUnicodeCharacter{2212}{-}

\makeatletter
    \renewcommand\paragraph{\@startsection{paragraph}{4}{\z@}%
    {2.25ex \@plus1ex \@minus.2ex}
    {-1em}%
    {\normalfont\normalsize\bfseries}}
\makeatother

\DeclareUnicodeCharacter{2212}{-}

\newcommand{\paramtheta}{\boldsymbol{\theta}}


\newcommand{\computerfont}[1]{{\fontfamily{cmtt}\selectfont #1} }

\newcommand{\xbs}{\boldsymbol{x}}
\newcommand{\ybs}{\boldsymbol{y}}
\newcommand{\thbs}{\boldsymbol{\theta}}

\renewcommand{\geq}{\geqslant}

\renewcommand{\th}{\theta } 

\def\orcid#1{\kern -0.4em\href{https://orcid.org/#1}{\includegraphics[keepaspectratio,width=0.7em]{images/orcid_logo.pdf}}}

\long\def\ca#1\cb{} 
\newcolumntype{P}[1]{>{\centering\arraybackslash}p{#1}}

\begin{document}
\title{A Continuous Variable Born Machine}
%
%
\author{Ieva \v{C}epait\.{e}\inst{1, 3} \and
Brian Coyle\inst{2} \and
Elham Kashefi\inst{1,2}}
\authorrunning{I. \v{C}epait\.{e} et al.}
%
\institute{School of Informatics, 10 Crichton Street, University of Edinburgh, United Kingdom, \and
CNRS, LIP6, Sorbonne Universit\'{e}, 4 place Jussieu, 75005 Paris, France, \\ \and
Department of Physics and SUPA, University of Strathclyde, Glasgow G4 0NG, United Kingdom.\\ }
\maketitle              
\begin{abstract}
Generative Modelling has become a promising use case for near term quantum computers. In particular, due to the fundamentally probabilistic nature of quantum mechanics, quantum computers naturally model and learn probability distributions, perhaps more efficiently than can be achieved classically. The Born machine is an example of such a model, easily implemented on near term quantum computers. However, in its original form, the Born machine only naturally represents \emph{discrete} distributions. Since probability distributions of a continuous nature are commonplace in the world, it is essential to have a model which can efficiently represent them. Some proposals have been made in the literature to supplement the discrete Born machine with extra features to more easily learn continuous distributions, however, all invariably increase the resources required to some extent. In this work, we present the \emph{continuous variable} Born machine, built on the alternative architecture of continuous variable quantum computing, which is much more suitable for modelling such distributions in a resource-minimal way. We provide numerical results indicating the models ability to learn both quantum and classical continuous distributions, including in the presence of noise.
\keywords{Continuous Variable\and Born Machine\and Generative Modelling}
\end{abstract}

\section{Introduction} \label{sec:intro}

With the dawn of the noisy intermediate-scale quantum (NISQ)~\cite{preskill_quantum_2018} device era comes a possibility of performing useful and large-scale computations that implement quantum information processing. While NISQ technologies do not entail fault-tolerance or large numbers of qubits (generally in the range of about 50-200) which we expect to be necessary for obtaining useful processing power, they do provide avenues for brand new methods of exploiting quantum information. One type of framework which can utilise the restricted architectures of NISQ devices is that of hybrid quantum-classical (HQC) methods, which have found many applications recently in fields like quantum chemistry \cite{peruzzo_variational_2014} and quantum machine learning (QML) \cite{benedetti_parameterized_2019}. These depend on dividing an algorithm into several parts which can be delegated to either quantum and classical servers, reducing the amount of quantum resources required to generate a solution.

In the field of quantum machine learning (QML), the benefits of HQC are key in approaches that employ paramterized quantum circuits (PQC) (also referred to as a quantum neural networks), which act as an Ansatz solution to some particular problem that can be optimised classically. QML has employed PQC's for several problems, including classification, \cite{farhi_classification_2018, schuld_quantum_2019, havlicek_supervised_2019, schuld_circuit-centric_2018, larose_robust_2020}, generative modelling, \cite{benedetti_generative_2019, liu_differentiable_2018, verdon_quantum_2017, romero_variational_2019, zoufal_quantum_2019}, and problems in quantum information and computation themselves \cite{morales_variational_2018, cincio_learning_2018, khatri_quantum-assisted_2019, cerezo_variational_2020, larose_variational_2019, romero_quantum_2017}. The `learnability' and expressive power of these models has been also been studied \cite{schuld_effect_2020, gil_vidal_input_2020, larose_robust_2020, coyle_born_2020, du_learnability_2020}. One can, for example, use quantum states as sources of probability distributions, with measurement playing the role of random sampling. These quantum states are obtained via PQC's which are composed of a number of tunable quantum gates with parameters that can be optimised using a classical subroutine. The process is typically iterative, requiring smaller quantum circuits with less depth to be run several times, thus decreasing information loss and quantum resource requirements. Alternatively, one could use coherent \emph{quantum} training procedures \cite{verdon_universal_2018} which may in fact be necessary to see quantum advantages in certain cases \cite{wright_capacity_2020}. 

QML is not restricted to a specific quantum computing paradigm and several models have shown potential in recent years, often divided into two broad categories: discrete- and continuous-variable \cite{lloyd_quantum_1999} (DV and CV) systems. DV systems allow for individually addressable states that are finite, often preferred for their analogous nature to classical computers. CV systems, on the other hand, deal with quantum states which behave as bosonic quantum modes (and are therefore referred to commonly as \textit{qumodes}) which effectively have infinite eigenstates with an added difficulty in addressing each individual state. This challenge notwithstanding, CV quantum computers allow for a far more effective manner of dealing with problems which require continuous values, making them a perfect candidate for modelling continuous distributions. Furthermore, much progress has been made in the field of QML using continuous variables, with software packages specifically created to deal with such scenarios~\cite{killoran_strawberry_2018}.

In this work, we use the CV model to study the \textit{Born machine} (BM) \cite{cheng_information_2018, benedetti_generative_2019, liu_differentiable_2018-1}, a mathematical model which generates statistics from a probability distribution $p(\xbs)$ according to the fundamental randomness of quantum mechanics, i.e.\@ Born's measurement rule (see Section \ref{sec:cv_born_machine_model}). We can generate samples of a distribution defined according to Born's rule via the measurement of some quantum state, making this a \textit{generative} QML method. Most commonly, a \emph{quantum circuit} Born machine (QCBM) is implemented, meaning the quantum state is prepared as by a PQC, although other definitions are possible, for example the parameterisation of density matrices via a combination of classical and quantum resources \cite{verdon_quantum_2019, liu_solving_2020,  martyn_product_2019}. The output distribution is then altered via a classical optimisation of the PQC's parameters in order to match the distribution of some target data. The target distribution may be the output of a quantum system - such as a quantum computer or an experimental measurement - making the quantum generative model naturally suited for learning it. Furthermore, Born machines are promising candidates for models which could demonstrate a quantum advantage in machine learning in the near term, \cite{du_expressive_2020, coyle_born_2020, alcazar_classical_2020, sweke_quantum_2020, liu_solving_2020}.

\section{Continuous Variable Quantum Computing}\label{sec:cv}

Here we give a brief overview of the nature of CV quantum computing as is pertinent to the rest of the work. For a full treatment of the topic, please refer to \cite{weedbrook_gaussian_2012, braunstein_quantum_2005}.
CV states can be represented in both phase space and Fock space  \cite{moyal_quantum_1949, goldstein_classical_2002, curtright_quantum_2011}, owing to the wave-particle duality of quantum mechanics. Importantly, both formulations give equivalent predictions about the behaviour of quantum systems. In Fock space, states $\{\ket{n}\}^{\infty}_{n=0}$ are eigenstates of the photon number operator $\hat{n}=\hat{a}^{\dagger}\hat{a}$ (where $\hat{a}$ and $\hat{a}^{\dagger}$ are the canonical creation and annihilation operators), giving $\hat{n}\ket{n} = n\ket{n}$. Importantly, the Fock basis is discrete as in the case of qubits (although with an infinite basis), so if we want to extract the benefits of continuous variables from CV systems, we need to consider them via the phase space approach.

In the phase space formulation, any state of a single qumode can be represented as a real-valued function $\textbf{F}(x,p)$, $(x,p) \in \mathbb{R}^2$ in phase space called the \textit{Wigner quasi-probability function}~\cite{wigner_quantum_1932, killoran_continuous-variable_2019, curtright_quantum_2011}. The two axes of phase space are then the quadrature variables governed by quadrature operators $\hat{x}$ and $\hat{p}$ which have a continuous and infinite basis with eigenstates $\ket{x}$ and $\ket{p}$, representing conjugate variables of a quantum system. The marginals of the Wigner function are the probability distributions of each of the quadrature variables, $|\psi(x)|^{2}$ and $|\phi(p)|^{2}$ where $\psi(x)$, $\phi(p)$ are complex-valued functions that can be understood as the amplitudes of their wavefunctions $\ket{x}$ and $\ket{p}$. 

The most basic qumode state is the \textit{vacuum} state:
\begin{align}\label{eqn:vacuum_state}
    \ket{0} = \frac{1}{\sqrt[4]{\pi\hbar}}\int dx  \,e^{\flatfrac{-x^2}{2\hbar}}\ket{x},
\end{align}
where $\ket{0}$ is its vector representation in the Fock basis ($n=0$). The vacuum state often serves as a starting point for a computation (similar to the discrete variable case) which can then be expressed as an evolution according to some bosonic Hamiltonian $H$ for a time $t$:
\begin{align}\label{eqn:hamiltonian_evolution}
    \ket{\Psi(t)} = e^{-i t H} \ket{0}
\end{align}
In the above, $H$ can be seen as a polynomial function of the quadrature operators $H = H(\hat{x}, \hat{p})$ with arbitrary but fixed degree. This Hamiltonian can be decomposed into a series of CV quantum gates (see Table \ref{table:gates}) in order to be implemented in a quantum circuit setting.

States $\ket{\Psi(t)}$ for which the Hamiltonian is at most quadratic in $\hat{x}$ and $\hat{p}$ are called \textit{Gaussian} \cite{weedbrook_gaussian_2012}. For the single qumode, the most common Gaussian quantum gates are \textit{rotation}: $R(\phi)$, \textit{displacement}: $D(\alpha)$ and \textit{squeezing}: $S(r,\phi)$ (see Table \ref{table:gates}). The simplest two-mode gate is the \textit{beamsplitter}, $BS(\theta)$, which is a rotation between two qumodes. These gates are parametrised accordingly: $\phi,\theta \in [0, 2\pi]$, $\alpha \in \mathbb{C} \cong \mathbb{R}^{2}$ and $r \in \mathbb{R}$.

Importantly, Gaussian gates are linear when acting on quantum states in phase space. On $n$ qumodes, a general Gaussian operator has the effect \cite{killoran_continuous-variable_2019}:
\begin{align}\label{eqn:symplectic_matrix_transformation}
    \mqty[\boldsymbol{x} \\ \boldsymbol{p}] \mapsto M \mqty[\boldsymbol{x} \\ \boldsymbol{p}] + \boldsymbol{\beta}
\end{align}

with symplectic matrix $M$~\cite{weedbrook_gaussian_2012} and complex vector $\boldsymbol{\beta} \in \mathbb{C}^{2n}$. The variables $\boldsymbol{x}$ and $\boldsymbol{p}$ contain the position and momentum information of each qumode, and are vectors in $\mathbb{C}^n$. Because of this linearity, Gaussian states and their operators are so-called ``easy" operations of a CV quantum computer. They can be efficiently simulated using classical methods and are generally easy to implement experimentally \cite{weedbrook_gaussian_2012, killoran_continuous-variable_2019}.

In order to achieve a notion of universality \cite{lloyd_quantum_1999} in the CV architecture we need the ability to construct a Hamiltonian as in \eqref{eqn:hamiltonian_evolution} that can translate to every possible state in the Hilbert space. This is done by introducing something called a \textit{non-Gaussian} transformation into the toolbox, which is essentially a non-linear transformation on $(x,p)$. Non-Gaussian gates involve quadrature operators that have a degree of 3 or higher. Some examples we consider in this work are the \textit{cubic phase gate} $V(\gamma)$ and the \textit{Kerr interaction} $K(\kappa)$ with $\gamma, \kappa \in \mathbb{C} \cong \mathbb{R}^{2}$. Non-Gaussian gates are generally difficult to implement and difficult to accurately simulate using a classical processor, as they cannot be decomposed in the same fashion as Gaussian gates  \eqref{eqn:symplectic_matrix_transformation}. This lack of decomposition implies that a completely accurate classical simulation of non-Gaussianity requires access to infinite matrices, thus requiring a choice of cut-off dimension which introduces some errors. On a quantum processor, this notion of infinity is inherent in the native physics of the hardware.

A list of the CV gates used throughout this work as well as their exponential operator form is presented in the Table \ref{table:gates} below. Note that the squeezing gate is parameterised by $\zeta$ and encompasses the parameters $r$ (squeezing strength) and $\phi$ (squeezing direction) via the relationship $\zeta = re^{i \phi}$:
\begin{table}[H]
\centering
\captionsetup{width=8cm, justification=justified, skip=5pt, belowskip=-10pt, font=small, labelfont={bf}}
\caption{A table of the most common Gaussian and several non-Gaussian transformations in the CV quantum computing model.}
\begin{tabu}{p{3cm}p{4cm}}
\textbf{Gate} & \textbf{Operators} \\[5pt]
\hline \\ [-1.5ex]
$R(\phi)$ & $\exp\left[ i \phi \hat{n} \right]$ \\[5pt]
$D(\alpha)$ & $\exp\left[\alpha\hat{a}^\dagger - \alpha^*\hat{a}\right]$ \\[5pt]
$S(\zeta)$ & $\exp\left[\frac{1}{2}\left(\zeta^*\hat{a}^2 - \zeta\hat{a}^{\dagger 2}\right)\right]$ \\[5pt]
$BS(\theta)$ & $\exp\left[\theta\left(\hat{a}^{\dagger}\hat{a} - \hat{a}\hat{a}^{\dagger}\right)\right]$ \\[5pt]
$V(\gamma)$ & $\exp\left[i\frac{\gamma}{6}\hat{x}^3\right]$ \\[5pt]
$K(\kappa)$ & $\exp\left[i\kappa\hat{n}^2\right]$ \\
\hline
\end{tabu}
\label{table:gates}
\end{table}

Finally, classical information can be extracted from CV systems via three types of measurement: Homodyne, Heterodyne and photon counting. For the purposes of the CVBM, we are interested largely in Homodyne detection, which returns real, continuous values of the two quadratures of a CV state. We discuss the measurement in more detail in the next section.

\section{A Continuous Variable Born Machine}\label{sec:cv_born_machine_model}

Having defined the CV model we can now construct a continuous variable \emph{Born machine} (CVBM). As discussed in the introduction, the Born machine can generate statistics sampled from a probability distribution according to Born's measurement rule:
\begin{align}\label{bornrule}
    p(\xbs) = |\braket{\xbs}{\psi(\paramtheta)}|^2
\end{align}
The state $\ket{\psi(\boldsymbol{\theta})}$ is generated by evolving the vacuum state \eqref{eqn:vacuum_state} according to a Hamiltonian $H$ that is constructed from CV gates given in Table \ref{table:gates}. These gates form a PQC which is parameterised by the variables governing each gate (indicated in \eqref{bornrule} by $\boldsymbol{\theta}$). These parameters should be easily tunable and allow for an evolution to any state that can serve as the solution to the given problem. 

Thus, in ML terms we can think of the state $\ket{\psi(\boldsymbol{\theta})}$ as a \textit{generative model}, which when measured in some pre-determined basis will generate samples of a distribution that is of interest. This model is parameterised by $\boldsymbol{\theta}$, which defines an $n$-qumode quantum circuit $U(\boldsymbol{\theta})$ made up of a set of quantum gates such that:
\begin{equation}\label{eqn:borncircuit}
    \ket{\psi(\paramtheta)} = U(\paramtheta)\ket{0}^{\otimes n}
\end{equation}
Now the final ingredient is the measurement of this parameterised state to extract the samples. This is key to the resource efficiency of the model. Here, a sample from a single qumode (either its position, $x$, or momentum, $p$) is extracted by measuring the corresponding quadrature using the infinite basis of operators in a homodyne measurement\footnote{In contrast, a \emph{heterodyne} measurement allows one to extract both position and momentum simultaneously from the state (with a failure probability given by the uncertainty principle) via the operators $\ketbra{x + ip}{x + ip}$ in a `coherent' basis.}~\cite{braunstein_quantum_2005}:
\begin{align}\label{eqn:homodyne_measurement}
\ketbra{\cos(\phi)\hat{x} + \sin(\phi)\hat{p}}{\cos(\phi)\hat{x} + \sin(\phi)\hat{p}}
\end{align}

\noindent By tuning the angle to $\phi=0$, we can extract the position, $x$, and for $\phi = \pi/2$, we can get the momentum $p$. For this work, we set $\phi = 0$ for all examples, and leave incorporation of alternative measurement strategies to future work. A full output sample, $\boldsymbol{x}$, can then be determined by measuring all $n$ qumodes.

The parameters of these gates then need to be \textit{trained} according to the problem at hand (as will be discussed in more detail in the next section). This generally involves a so-called \textit{cost function} whose value depends on the output of the model as well as some training data samples and which needs to be minimised by varying the values of the parameters of the circuit $U(\boldsymbol{\theta})$. This is referred to as \emph{empirical risk minimisation} \cite{vapnik_principles_1992} in the classical ML literature. The choice of cost function and gates with which the circuit is built is problem-dependent, although there are often multiple cost functions which can be utilised for the same model, varying in efficiency and accuracy.

Once the value of the cost function converges to a minimum, the CVBM should be able to generate samples that behave like those of a \emph{target} distribution (to be learned). Importantly, the purpose of the CVBM is not to exactly reproduce the samples that were fed into it during the training, but rather to emulate the original distribution from which they arose. This makes the CVBM a synthetic data generator if the original is no longer available, as well as opening up new avenues to study the model itself: the number of, and importance of, the degrees of freedom of the model, for example (given by the gates and their influence on the final distribution). It also means that there is not necessarily a requirement to reach a global minimum of the cost function once training is completed, as a local minimum may lead to a model that is just as effective in achieving the desired samples. Furthermore, different cost function minima might present further insight into the target distribution and its behaviour through the parameters of the model. This is particularly pertinent when talking about QML methods versus classical ones, as the model itself is inherently quantum and the physics may be studied natively within the experimental set-up once the model is trained.

The notion of `quantum' parameters leads to one of the key reasons to employ a CVBM in learning continuous probability distributions which are quantum mechanical in nature: as well as lending itself to further analysis once trained, it also promises to be far more efficient and accurate in learning a quantum state than any classical model. Primarily this is down to the fact that the degrees of freedom we would expect to affect the distributions are captured explicitly by the parameters $\boldsymbol{\theta}$, saving us gymnastics of developing a new mathematical model that might strive to capture their behaviour.  While it is possible to write emulators of quantum systems on classical machines for simple cases, once we are outside of the regime of being able to simulate a quantum system classically, the Born Machine can act as an exponentially complex black-box used to generate samples that could otherwise not be obtained, governed by a tractable number of parameters that can be understood and manipulated. This is irrespective of whether or not we know anything about the quantum system in question, which makes it an advantage even over classical methods that may employ approximations and other shortcuts in dealing with the exponential size of the Hilbert space they need to explore. 

Finally, we may add that while a CVBM may be perfectly suited for learning continuous probability distributions which are quantum mechanical in nature, there is no reason to believe that they could not be used to learn classical distributions. Indeed, a Gaussian distribution \eqref{eqn:gaussian_pdf} is parameterised in a way that has direct correspondence with the Squeezing $S(\zeta)$ and Displacement $D(\alpha)$ gates (see Table \ref{table:gates}) via its standard deviation and mean respectively. Thus, a CVBM circuit composed of only those two gates is expressive enough to capture any Gaussian distribution (for multidimensional Gaussians we simply need to add a qumode for each new dimension). Other classical distributions are not necessarily as straightforward, but we expect that a well-tailored CVBM (size, gate-set and their sequence may all be important factors) could be suitable adapted.

\subsection{Previous Work}\label{sec:previous_work}
The CVBM is introduced as a resource efficient method of generating continuous probability distributions. In order to do so, it is instructive to revisit other attempts to generate continuous distributions using quantum generators (here we compare only the sample generation mechanism, and not specifics of the model or training, which we discuss for the CVBM in the following sections). As discussed above, the CVBM allows sample generation via single homodyne measurements of a particular quadrature, for example the position $\boldsymbol{x}$ \footnote{Note that the continuous values extracted are actually approximations to the true quadratures, since they are projected onto squeezed states~\cite{killoran_strawberry_2018}}, so an element of a real sample vector is generated without any post-processing.

Firstly, the work of~\cite{liu_differentiable_2018} numerically tested the performance of a QCBM when trained differentiably to learn a continuous distribution composed of a mixture of Gaussian distributions. This work used $10$ qubits which results in approximating the real distribution using $2^{10} = 1024$ basis states without any measurement post-processing. Secondly, the work of \cite{romero_variational_2019, anand_experimental_2020} explicitly addresses this question of generating real valued distributions using a Born machine as a generator in an adversarially trained scenario. Specifically, previous works utilised a Born machine to generate $n$-bit binary strings, $\boldsymbol{x} \in \{0, 1\}^n$ by simply taking the measurement result from measuring (for example) every qubit in the computational basis, which generates one bit, $x_i \in \{0, 1\}$ per qubit. In contrast, the innovation of \cite{romero_variational_2019} was to instead evaluate the \text{expectation} value (for example) of the Pauli-$Z$ observable\footnote{Note that \cite{romero_variational_2019} allows measurements of any single qubit observable, $P$, which does not have to be the Pauli $Z$ operator. However, for illustration purposes, we assume the measurement is done in the computational basis.} from the measurement results to generate a real value, $x_i \in [-1, 1]$ for each qubit. 
More concretely, the final sample is an $n$ bit string, $\boldsymbol{x} \in \mathbb{R}^n$ generated by the following process:
\begin{equation}\label{eqn:romero_continuous}
    \{(x^m_1, x^m_2, \dots, x^m_n)\}_{m=1}^M  \rightarrow \tilde{\boldsymbol{x}} := (\langle Z_1 \rangle, \langle Z_2 \rangle, \dots, \langle Z_n \rangle) \rightarrow f(\tilde{\boldsymbol{x}}, \boldsymbol{\phi}) := \boldsymbol{x}
\end{equation}
The intermediate quantity, $\tilde{\boldsymbol{x}}$ is the vector of expectation values for each qubit, $\tilde{\boldsymbol{x}} \in [-1, 1]^n$, which is fed into a classical function, $f:\mathbb{R}^n \rightarrow \mathbb{R}^n$. The function could be, for example,
one layer of a \emph{classical} feedforward neural network, where $\boldsymbol{\phi} := (\boldsymbol{W}, \boldsymbol{b})$ are the weights and biases of the network.
This method addresses the continuous distribution problem, but at the expense of adding $\mathcal{O}(n/\epsilon^2\log\delta)$ extra measurements which must be evaluated (and hence circuits which must be run) in order to compute the expectation values, $\tilde{\boldsymbol{x}}$, with sufficiently high probability ($1-\delta$) by Hoeffding's inequality \cite{hoeffding_probability_1963}. Hence, this adds a large overhead to the efficiency of the model in order to generate a single sample,  $\boldsymbol{x} \in \mathbb{R}^n$. 
An alternate approach, intermediate to \cite{liu_differentiable_2018} (using a completely discrete output) and \cite{romero_variational_2019}, (which increases number of circuit evaluations), is to use a QCBM, with $n$ qubits, and convert the resulting $n$ bit binary outputs into real valued numbers with a corresponding precision. Again, this method is resource intensive as $n$ qubits are required to generate each vector element in $\mathbb{R}$, and so is less ideal than our main approach.
We illustrate this latter method numerically in Section \ref{sec:numerical_results} to contrast with the efficiency of the CVBM to learn a Gaussian distribution.

\section{Training}\label{sec:training}

In this work, we are specifically interested in training a CVBM to generate data samples which behave as if they were sampled from some unknown target continuous probability distribution. We need to be able to do this while having access only to some limited number of training samples from the distribution we wish to learn as well as samples from the CVBM at any point in the training process. The metric we choose to implement in training our model based on these requirements is the \textit{Maximum Mean Discrepancy} (MMD), which we discuss in the next section.

\subsection{Maximum Mean Discrepancy}

The MMD is a suitable metric for our purposes in several respects, having been used to train a discrete Born machine in the past \cite{liu_differentiable_2018} and requiring a relatively low number of samples from each distribution \cite{sriperumbudur_integral_2009}. 

A key component of the MMD is the \textit{kernel function} (defined below) and ML methods which use them are unsurprisingly known collectively as `kernel methods'\cite{gretton_kernel_2007, hofmann_kernel_2008, schuld_quantum_2019}. The so-called `kernel-trick' is useful for comparing data points, even when the underlying feature map may be difficult to compute. They use a similarity measure $\kappa(\boldsymbol{x}, \boldsymbol{x}')$ between two data points $\boldsymbol{x}$ and $\boldsymbol{x}'$ in order to construct models that capture the properties and patterns of a data distribution. This measure of distance is related to the inner products of a \textit{feature space}, the idea of which is key to kernel methods.

Kernels are symmetric functions of the form $\kappa : \mathcal{H} \times \mathcal{H} \rightarrow \mathbb{C}$ wherein $\mathcal{H}$ is a Hilbert space (the aforementioned feature space). The main idea behind Kernel Methods is to embed data samples from their original sample space $\mathcal{X}$ into a space $\mathcal{H}$ via a mapping $\phi: \mathcal{X} \rightarrow \mathcal{H}$. This is called a \textit{feature map} and plays the the role of a `filter' for the samples, with the aim to, say, achieve a reduction in dimensionality or some form of useful restructuring of the data that might aid in the training procedure. A (positive definite, real-valued) kernel inner product should also have the property $\kappa(\boldsymbol{x} ,\boldsymbol{x}') \geq 0$ as well as being symmetric: $\kappa(\boldsymbol{x} ,\boldsymbol{x} ') = \kappa(\boldsymbol{x} ',\boldsymbol{x} )$.

A typical example is the so-called `Gaussian' kernel:
\begin{align}\label{eqn:gaussian_kernel}
    \kappa_G(\boldsymbol{x} , \boldsymbol{y} ) = e^{-\flatfrac{||\boldsymbol{x} -\boldsymbol{y} ||^2}{2\sigma^2}}
\end{align}

\noindent where $\boldsymbol{x}$ and $\boldsymbol{y}$ are two data points (generally called \textit{feature vectors}), $||\cdot||^2$ is the Euclidean distance between them and $\sigma$ is a constant which represents the variance or the `bandwidth' which determines the scale at which the points are compared. The kernel value decreases with distance between the two data points and as such is an effective similarity measure \cite{gretton_kernel_2012}.

\paragraph{Quantum Kernels: }\label{par:quantumkernel}

One nice consequence of implementing kernels is that any positive definite kernel can be replaced by another and this opens up the doors to a brand new approach of improving ML algorithms. Before moving on to the MMD itself, we note the fact that quantum states are like feature vectors themselves in that they also reside in Hilbert spaces and allow for a very natural definition of a \textit{quantum kernel}. if we find an effective way of encoding input data points $\boldsymbol{x} \in \mathcal{X}$ into quantum states $\ket{\phi(\boldsymbol{x})}$ then we end up with a feature map. Furthermore, the overlap of two quantum states can then be implemented as a kernel distance, with increasing orthogonality of two states leading to a decrease in kernel value. 
\begin{equation}\label{eqn:quantum_kernel}
    \kappa(\boldsymbol{x} ,\boldsymbol{x} ') = \bra{\phi(\boldsymbol{x} )}\ket{\phi(\boldsymbol{x} ')}
\end{equation}
In order to compute this overlap, one could use CV SWAP-test like primitives, \cite{chabaud_optimal_2018, kumar_optimal_2020}, but due to the special form of the kernel, it can be evaluated more simply on NISQ devices \cite{havlicek_supervised_2019} by running a unitary and then its inverse. Quantum kernels have already been explored within the CVQC framework and we refer to \cite{schuld_quantum_2019} for a full treatment of the topic. For the purposes of this work, we employ several different encodings of classical data samples into CV quantum kernels to explore their efficiency and effectiveness in training the CVBM. While determining a good kernel mapping is a complex and generally problem-dependent issue, there are grounds for the use of quantum kernels as they may promise a far more complex mapping than anything that can be achieved classically and could prove to be useful in settings involving large data sets and particularly in terms of high dimensionality or correlation within the data. Quantum kernels were first investigated in the context of generative modelling in \cite{coyle_born_2020, kubler_quantum_2019}.

\paragraph{MMD Estimator: }

The kernel acts as a feature map which embeds data samples in a \textit{Reproducing Kernel Hilbert Space} (RKHS) \cite{schuld_quantum_2019} and it can be shown that the MMD is a metric describing exactly the difference in mean embeddings between two distributions from which the samples are collected \cite{gretton_kernel_2007, gretton_kernel_2012}. With this, we can define a cost function associated with the metric for distributions $P$ and $Q$:
\begin{equation}\label{eqn:mmd_cost_function}
    \mathcal{L}_{\mathrm{MMD}}[P,Q] = \underset{\substack{\boldsymbol{x}\sim P, \\ \boldsymbol{y}\sim P}}{\mathbb{E}}(\kappa(\boldsymbol{x}, \boldsymbol{y})) + \underset{\substack{\boldsymbol{x}\sim Q,\\ \boldsymbol{y}\sim Q} }{\mathbb{E}}(\kappa(\boldsymbol{x}, \textbf{y})) - \underset{\substack{\boldsymbol{x}\sim P,\\ \boldsymbol{y}\sim Q} }{2\mathbb{E}}(\kappa(\boldsymbol{x}, \boldsymbol{y})), 
\end{equation}
where $\xbs \sim Q$ indicates a sample $\xbs$ drawn from distribution $Q$ and $\kappa$ is the MMD kernel. Given i.i.d. samples drawn from each distribution, the MMD can be estimated by replacing the expectation values in \eqref{eqn:mmd_cost_function} by their empirical values to produce the (unbiased) \textit{MMD estimator}:
\begin{equation}\label{eqn:mmd_estimator}
    \mathcal{L}_{\mathrm{MMD}}[P,Q] = \frac{1}{M(M-1)}\sum_{i\neq j}^M \kappa(\boldsymbol{x}_i, \boldsymbol{x} _j) + \frac{1}{N (N-1)}\sum_{i\neq j}^N \kappa( \boldsymbol{y}_i, \boldsymbol{y}_j) - \frac{2}{MN}\sum_{i,j}^{M, N} \kappa(\boldsymbol{x}_i,\boldsymbol{y}_j)
\end{equation}
with $M$ samples $\tilde{\xbs} := (\xbs_1, ... ,\xbs_M)$ drawn from distribution $P$ and $N$ samples $\tilde{\ybs} := (\ybs_1, ..., \ybs_N)$ drawn from $Q$. Given a large enough number of samples, \eqref{eqn:mmd_estimator} should converge to the true expectations given in \eqref{eqn:mmd_cost_function}. The MMD is useful as an estimator due to the relatively low number of samples it requires to satisfy the above requirement~\cite{sriperumbudur_integral_2009} ($\mathcal{O}(\epsilon^{-2})$ to reach a precision $\epsilon$). The estimator allows for a practical, numerical implementation to train a model in order to reproduce samples from a target distribution. In this work, we want to apply it to the CVBM.

\subsection{Training the CVBM} \label{ssec:training_cvbm}

The CVBM is parameterised by the operations given in Table \ref{table:gates} and in order to be able to train it efficiently, we need to determine a way to quickly navigate the parameter space of the MMD estimator for any quantum circuit composed of any of the given operations. A common method used in many machine learning algorithms is \textit{gradient descent}, often implemented in a stochastic fashion, with many varieties to facilitate a trade off between accuracy and speed~\cite{ruder_overview_2017}.

A key component of gradient descent is the calculation of loss function gradients with respect to each of the parameters. For each parameter $\theta_k$, we need to determine $\partial_{\theta_k} \mathcal{L}_{\mathrm{MMD}}[P(\thbs),Q]$ wherein the distribution $P(\thbs)$ is a CVBM circuit composed of gates parameterised by $\thbs := \th_1, ... , \th_l$.

The work of \cite{liu_differentiable_2018} shows that in the case of the discrete Born machine, by measuring an observable $\hat{O} = \ket{\boldsymbol{x}}\bra{\boldsymbol{x}}$, the gradient of the probability distribution generated by a QCBM, $p_{\th}$, with respect to parameter $\th_k$ is:
\begin{equation}\label{eqn:bm_prob_grad}
    \pdv{p_{\theta}(\boldsymbol{x})}{\theta_k} = \frac{1}{2}(p_{\theta_k^+}(\boldsymbol{x}) - p_{\theta_k^-}(\boldsymbol{x}))
\end{equation}
Where the parameters $\th_k^{\pm}$ imply that the gate parameter $\th_k$ has been shifted by an amount $\pm \frac{\pi}{2}$. In the CVQC case, this needs to be adapted for CV operators by adding specific scaling factors and choosing different shift amounts for $\th_k^{\pm}$ depending on the circuit gate (see Table \ref{table:gradients}). We chose to implement the analytic gradients derived in \cite{schuld_evaluating_2019} (also see \cite{mitarai_quantum_2018, crooks_gradients_2019, banchi_measuring_2020}) for the Gaussian gates of Table \ref{table:gates} as well as additional approximations of gradients for the cubic phase gate $V(\gamma)$ and Kerr gate $K(\kappa)$ (see Appendix \ref{app_a} for details of the approximation and a more detailed derivation of the gradients).

The gradient of the MMD Estimator with respect to the CVBM parameters can be described by:
\begin{equation}\label{eqn:mmd_grad}
    \pdv{\mathcal{L}_{\mathrm{MMD}}}{\theta_k} \approx \frac{1}{RM}\sum_{i,j}^{R, M} \kappa(\boldsymbol{a}_i, \boldsymbol{x}_j) - \frac{1}{SM}\sum_{i,j}^{S, M} \kappa(\boldsymbol{b}_i,\boldsymbol{x}_j) - \frac{1}{RN}\sum_{i,j}^{R, N} \kappa(\boldsymbol{a}_i, \boldsymbol{y}_j) + 
    \frac{1}{SN}\sum_{i,j}^{S, N} \kappa(\boldsymbol{b}_i, \boldsymbol{y}_j), 
\end{equation}
where $p, q$ samples $\tilde{\boldsymbol{a}} := \{\boldsymbol{a}_1,...,\boldsymbol{a}_R\}$, $\tilde{\boldsymbol{b}} := \{\boldsymbol{b}_1,...,\boldsymbol{b}_S\}$ are drawn from shifted circuits $p_{\theta_k^+}(\boldsymbol{a})$ and $p_{\theta_k^-}(\boldsymbol{b})$ respectively while $\tilde{\xbs} = \{\boldsymbol{x}_1, ... , \boldsymbol{x}_M\}$ and $\tilde{\ybs} = \{\boldsymbol{y}_1, ... ,\boldsymbol{y}_N\}$ are drawn from the CVBM and the target distribution. Using the above equation and the scaling factors and shift amounts given in Table \ref{table:gradients}, we can determine the gradient of the MMD Estimator with respect to any gate from the set given in Table \ref{table:gates}. Note that the error on the gradient for non-Gaussian gates scales approximately as the small-angle approximation error $\sinh(t) \approx t$.

\begin{table}[ht]
\centering
\captionsetup{width=12cm, justification=justified, skip=5pt, belowskip=-10pt, font=small,labelfont={bf}}
\caption{The gradients of the MMD cost function with respect to the parameters of each possible gate. The expression $\partial_{\theta_k} \mathcal{L}_{\mathrm{MMD}}$ is equivalent to \eqref{eqn:mmd_grad}. For the non-Gaussian gates $V(\gamma$) and $K(\kappa)$, the gradients are approximations for small enough parameter shifts $t$ rather than analytic gradients. It should be noted that while these latter approximations increase in accuracy with decreasing $t$, there is an inherent difficulty in implementing such small parameter shifts on current CV hardware. However, in practice it is a far more effective and accurate approach than any finite-difference method even with such drawbacks.
v}
\begin{tabu}{P{2cm} P{3cm} P{6.5cm}}
\textbf{Gate} & \textbf{Shift Amount} & \textbf{MMD Estimator Gradient}\\[5pt]
\hline \\ [-1.5ex]
$R(\phi)$ & $\phi \pm\frac{\pi}{2}$ & $\partial_{\phi} \mathcal{L}_{\mathrm{MMD}} = \partial_{\theta_k} \mathcal{L}_{\mathrm{MMD}}$ \\[8pt]
$D(\alpha)$ & $\alpha \pm s, s \in \mathbb{R}$  & $\partial_{\alpha} \mathcal{L}_{\mathrm{MMD}} = \frac{1}{s} \times \partial_{\theta_k} \mathcal{L}_{\mathrm{MMD}}$ \\[8pt]
$S(r,\phi)$ & $r \pm s, s \in \mathbb{R}$ &  $\partial_{r} \mathcal{L}_{\mathrm{MMD}} = \frac{1}{\sinh (s)} \times \partial_{\theta_k} \mathcal{L}_{\mathrm{MMD}}$ \\
$BS(\theta, \phi)$ & $\theta,\phi \pm\frac{\pi}{2}$ & $\partial_{\theta, \phi} \mathcal{L}_{\mathrm{MMD}} = \partial_{\theta_k} \mathcal{L}_{\mathrm{MMD}}$ \\
$V(\gamma)$ & $\gamma \pm t, t \ll 1$ &  $\partial_{\gamma} \mathcal{L}_{\mathrm{MMD}} = \frac{i}{t} \times \partial_{\theta_k} \mathcal{L}_{\mathrm{MMD}}$ \\ 
$K(\kappa)$ & $\kappa \pm t, t \ll 1$ & $\partial_{\kappa} \mathcal{L}_{\mathrm{MMD}} = \frac{i}{t} \times \partial_{\theta_k} \mathcal{L}_{\mathrm{MMD}}$ \\ 
\hline
\end{tabu}
\label{table:gradients}
\end{table}

We train the CVBM iteratively by implementing batch gradient descent for each parameter $\th_k$: 
\begin{align}\label{grad_desc}
    \theta^{(t+1)}_k = \theta^{t}_k - \mu\,  \partial_{\theta}\mathcal{L}_{\mathrm{MMD}},
\end{align}
with $t$ representing the iteration number and $\mu$ being the learning rate \cite{ruder_overview_2017}. This is done sequentially for each $\th_k$ per iteration of the training in order to update their values in a way that minimises the MMD estimator \eqref{eqn:mmd_estimator}. The training terminates once a set number of iterations has been performed or else when the cost function to observed to converge.

\begin{figure*}[ht]
    \centering
    \includegraphics[width=0.75\textwidth]{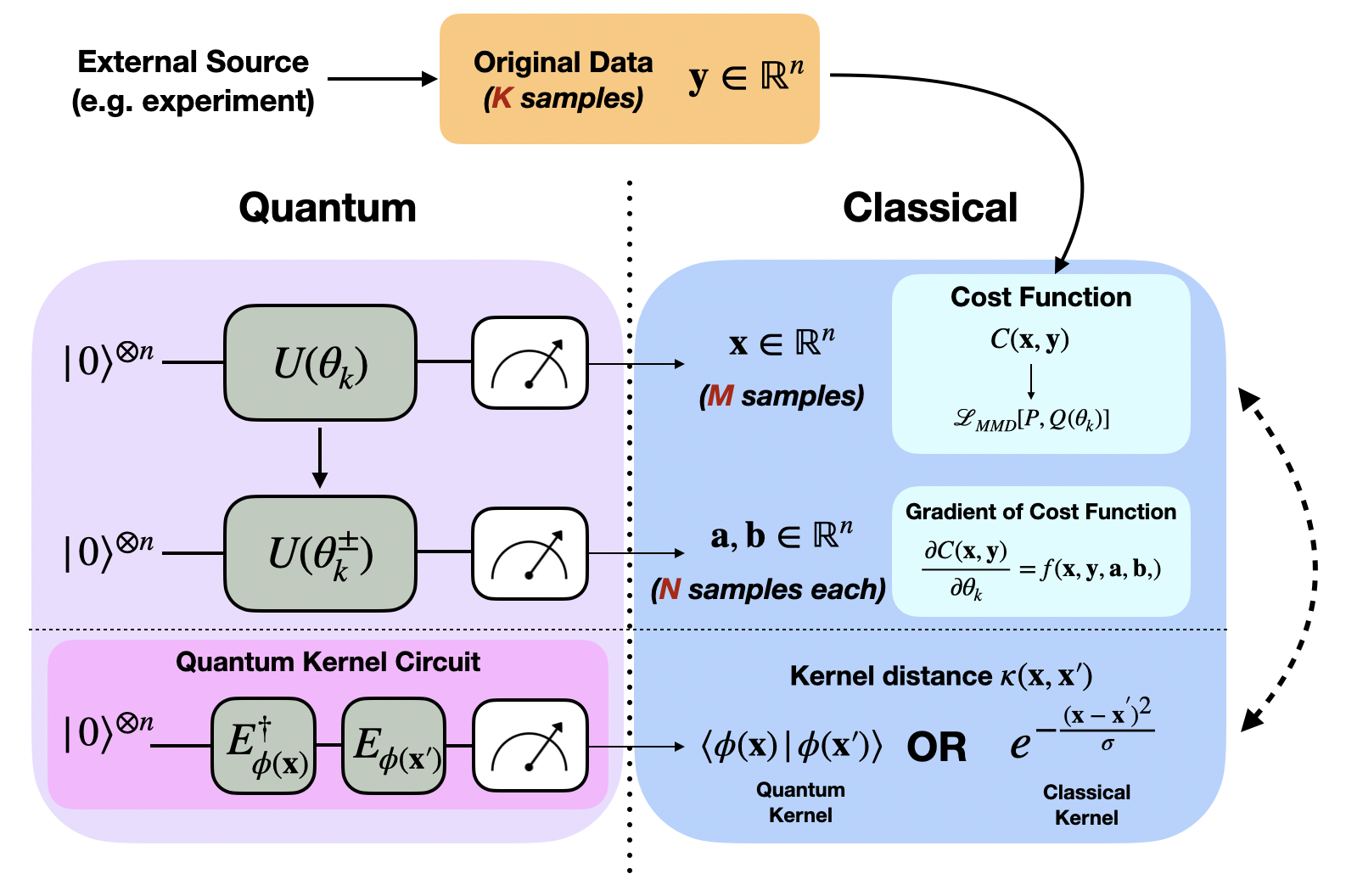}
    \captionsetup{width=\linewidth, justification=justified,font=small, belowskip=-10pt, labelfont={bf}}
    \caption{Components of the Hybrid Quantum-Classical CVBM. The Quantum hardware is used to produce measurement samples of the parameterised circuit $U(\theta_k)$ as well as samples from circuits with shifted parameter values $\theta_k^{\pm}$. These samples are then employed to classically compute the value of the cost function as well as its gradient in order to update the circuit parameters. If the cost function implements a kernel, this can be a classical function (such as the Gaussian kernel displayed in the Figure above) or quantum in nature, for example by running the corresponding encoding circuit $E_{\phi(\boldsymbol{x})}$, with $\phi(\boldsymbol{x})$ being a quantum feature encoding for a data point $\boldsymbol{x}$. }\label{fig:training_diagram}
\end{figure*}

In Figure~\ref{fig:training_diagram}, we emphasise the key ingredients of the model; a compact data encoding method for continuous distributions via the CVBM itself, an efficient training method via the MMD and a potentially classically-hard-to-compute ingredient in the quantum kernel. In the next section we validate its efficacy through numerical results on example classical and quantum distributions.

\section{Numerical Experiments}\label{sec:numerical_results}

Here we present numerical results demonstrating the performance of the model on both classical and quantum data sets, as well as the impact of a noise channel and several different kernels. The simulations are implemented using Xanadu's Strawberry Fields API \cite{killoran_strawberry_2018}, which uses the symplectic matrix approach in simulating Gaussian states (see \eqref{eqn:symplectic_matrix_transformation}) and a truncation of dimensions in Fock space when dealing with non-Gaussianity. Throughout all experiments, a cut-off dimension of 7 was used unless explicitly stated otherwise, owing to the fact that states with higher Fock numbers are likely to have little impact on the statistics of the single and two-qumode states that were used.

\begin{figure}[ht]
    \centering
    \includegraphics[width=\columnwidth]{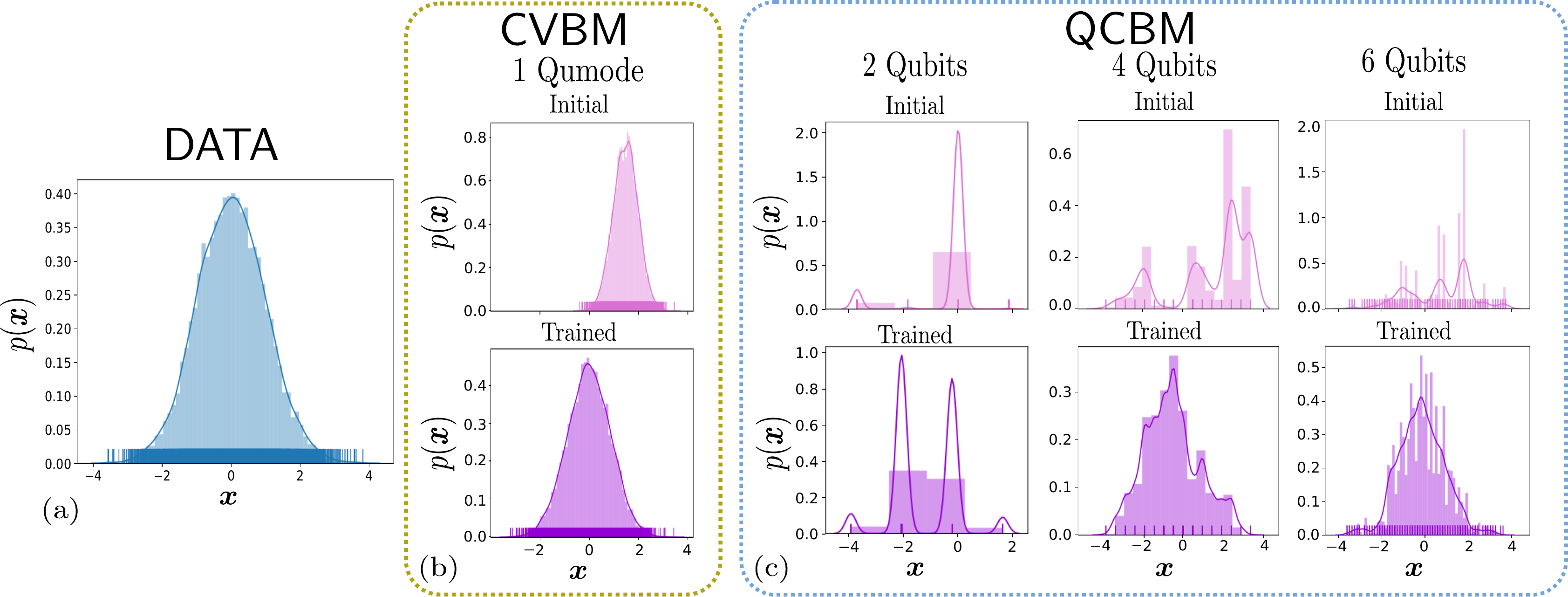}
    \captionsetup{width=\linewidth, justification=justified,font=small, belowskip=-10pt, labelfont={bf}}
    \caption{Comparing a QCBM to our CVBM for learning a simple Gaussian distribution (a) with mean, $\mu=0$, and standard deviation $\sigma = 1$. In (c), we show the QCBM with increasing numbers of qubits ($2, 4$ and $6$ qubits respectively, which results in increased precision for each sample. The QCBM is trained with the Adam optimiser \cite{kingma_adam_2015} using the Sinkhorn divergence~\cite{coyle_born_2020, feydy_interpolating_2018, genevay_learning_2017, genevay_sample_2018}. In (b), we show a single qumode with one parameterised squeezing and displacement gate for the CVBM, which can produce a much better fit to the data distribution with significantly fewer resources.}
    \label{fig:cvbm_vs_dvbm_gaussian_dist}
\end{figure}

\subsection{A Classical Distribution}
\label{ssec:numerical_classical_distributions}

An obvious choice of classical distribution to train the model on is the canonical Gaussian distribution. The classical Gaussian probability density function (PDF) is parameterised by mean $\mu$ and standard deviation $\sigma$, (which can correspond to the displacement $\alpha$ and squeezing $\zeta$ operators as discussed above). To generate data representing the (single mode) classical Gaussian PDF, we take M samples from $\pi$, as given by:
\begin{equation}\label{eqn:gaussian_pdf}
    y \sim \pi(y) = \mathcal{N}(\mu, \sigma^2) = \frac{1}{ \sqrt{ 2 \pi \sigma^2 } }e^{-\frac{1}{2}\left(\frac{y-\mu}{\sigma}\right)^2} , 
\end{equation}

To illustrate the advantage of using our method with a natural continuous encoding, we compare to a discrete variable Born machine, which outputs binary strings of length $n$. However, as mentioned above, in order to use a discrete Born machine to learn real valued distributions, we must decide on the precision we wish to use. In \figref{fig:cvbm_vs_dvbm_gaussian_dist} we use a discrete Born machine (QCBM) with $2, 4$ and $6$ qubits to learn a Gaussian distribution, $\mathcal{N}(0, 1)$ with increasingly higher precision, and also we use the CVBM with a single qumode to learn the same distribution. For the QCBM, we simulate the results using Pyquil~\cite{smith_practical_2016} and for an Ansatz we choose a hardware efficient layered Ansatz. Each layer consists of $\mathsf{CZ}$ matching the topology of a sublattice of the Rigetti \computerfont{Aspen-7} chip, with parameterised $\mathsf{R}_y(\theta)$ gates. For the $2$, $4$ and $6$ qubit QCBMs, we use $8, 4$ and $6$ layers respectively resulting in $64, 16$ and $36$ trainable parameters. To convert between the binary outputs of the QCBM, and the continuous valued data, we use a simple conversion algorithm described in~\cite{kondratyev_market_2019}. We also mention that the comparisons we make here are preliminary, and open the door to rigorous comparison and benchmarking between the CVBM and other models.

\subsection{Quantum Distributions} \label{ssec:numerical_quantum_distributions}

\begin{figure}[ht]
     \centering
    \begin{subfigure}[t]{0.46\textwidth}
    \centering
     \begin{tikzpicture}
          \node (img)  {\includegraphics[width=\textwidth]{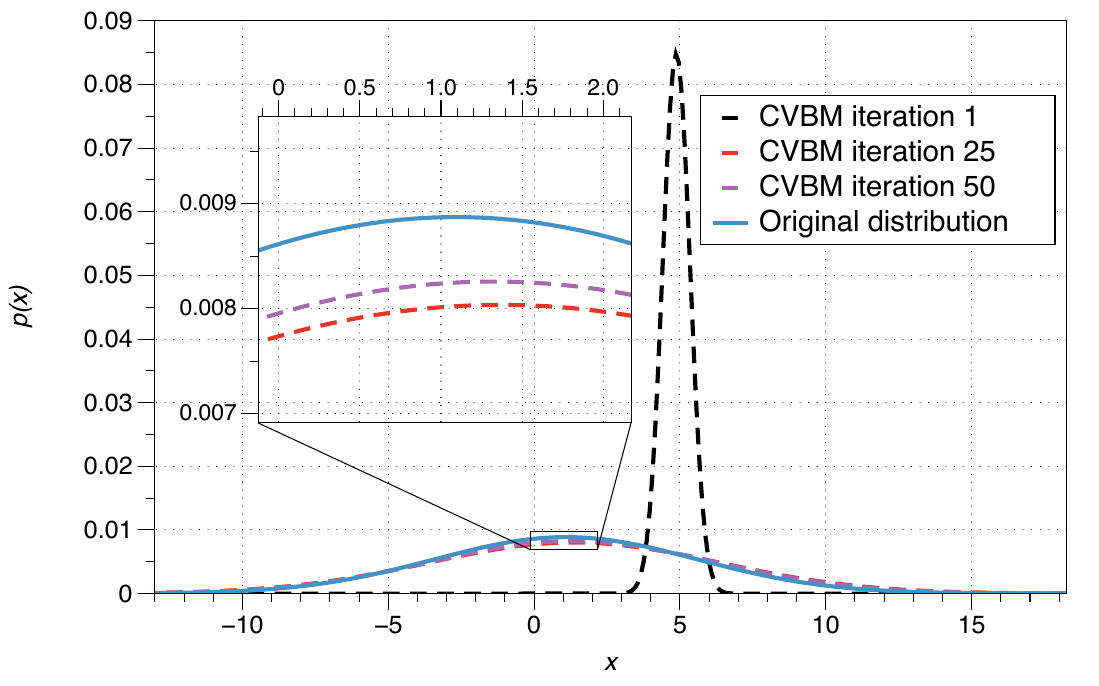}};
     \end{tikzpicture}
         \captionsetup{font=small, labelfont={bf}}
         \caption{Gaussian state distribution parameterised by $S(\zeta)$ and $D(\alpha)$.}
         \label{subfig:learning_gaussian_state}
    \end{subfigure}
     ~ 
 \begin{subfigure}[t]{0.46\textwidth}
 \centering
     \begin{tikzpicture}
         \node (img)  {\includegraphics[width=\textwidth]{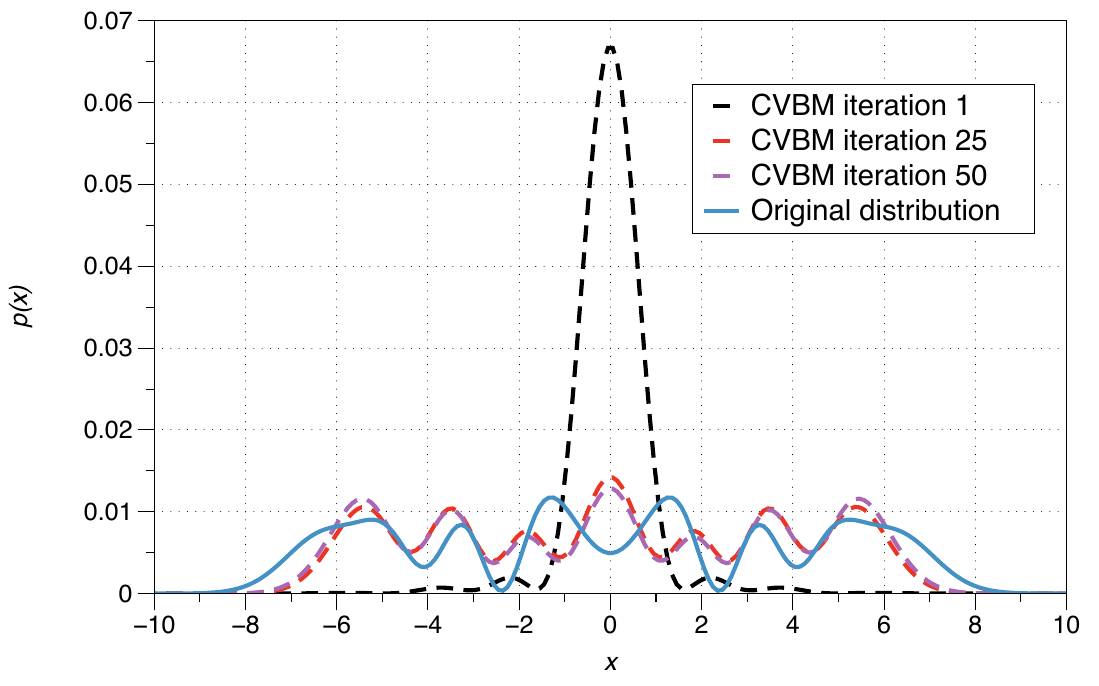}};
    \end{tikzpicture}
         \captionsetup{font=small, labelfont={bf}}
         \caption{Non-Gaussian state distribution parameterised by $S(\zeta)$ and $V(\gamma)$.}
         \label{subfig:learning_non_gaussian_gate}
     \end{subfigure}
     \captionsetup{justification=justified,font=small, belowskip=-10pt, labelfont={bf}}
     \caption{Snapshots of the CVBM learning process at iterations 1, 25 and 50 for two single-qumode quantum states. 50 data samples are used for the CVBM and target distributions respectively with 30 samples from shifted circuits for gradient computation.} \label{fig:iterations}
\end{figure}

Next, we focus on learning \emph{quantum} distributions, i.e.\@ one which arise as a result of measurements on a quantum state. This in some sense is equivalent to learning parameters which affect a quantum state, and in turn, gives some information about the parameters of the unitary which prepared the state, as a form of weak compilation, as noted by \cite{coyle_born_2020}. We generate data by sampling CV circuits with different parameters and then use a CVBM that is parameterised by the same gates. \figref{fig:iterations} shows the learning process for a Gaussian and a non-Gaussian state using a classical kernel (Eq. \ref{eqn:gaussian_kernel}). 

It should be noted that the slightly larger discrepancy in \ref{subfig:learning_non_gaussian_gate} is due to the difficulty in capturing the parametrisation and its relationship to the distribution of such states via sampling. This issue calls for a future investigation into finding better kernel functions for non-Gaussianity - ones which could capturing this behaviour more effectively and efficiently. In the Gaussian case, however (\ref{subfig:learning_gaussian_state}), the model converges to a very good approximation of the original distribution within less than 25 iterations of training while requiring only 50 samples (measurements) of the quantum state. This bodes well for the application of the method in - for example - characterising experimental data in the CV setting with very few data samples. 

\subsection{Quantum Kernels}

\begin{figure}[h!]
    \centering
    \includegraphics[width=0.55\linewidth]{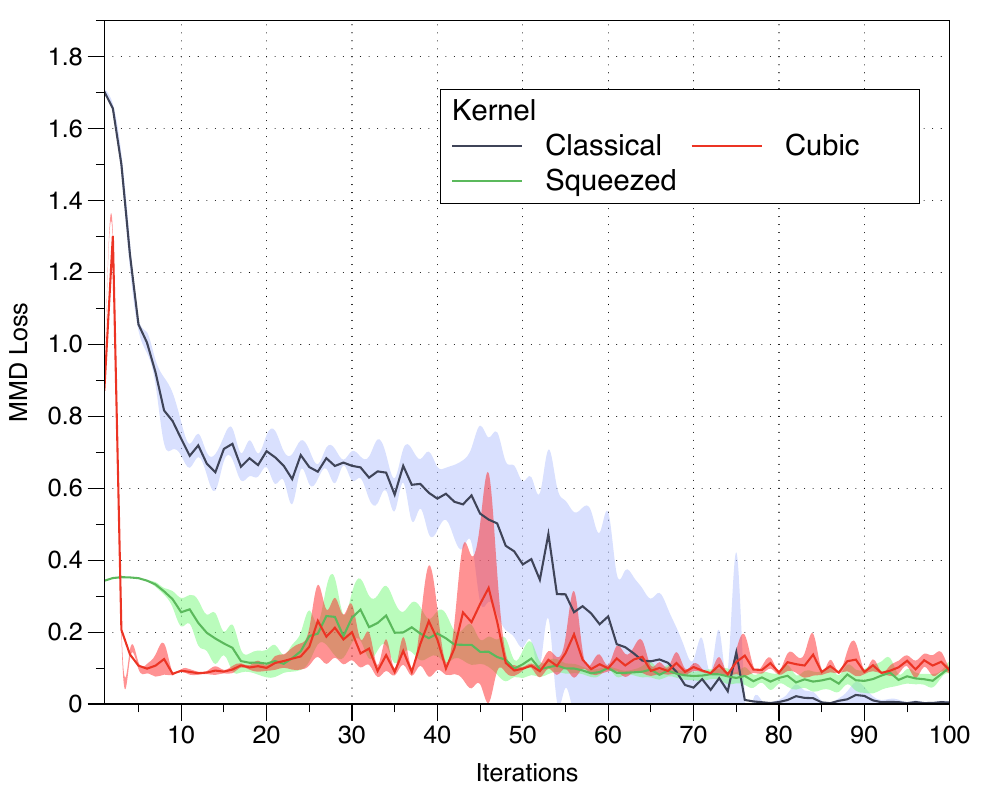}
    \captionsetup{width=\linewidth, justification=justified, font=small, belowskip=-10pt, labelfont={bf}}
    \caption{A 3 qumode Gaussian CVBM composed of squeezing, displacement and beamsplitter gates is trained to learn a distribution generated by another CV circuit made of the same components in a different order. The plot illustrates the behaviour of the loss when using a classical Gaussian kernel \eqref{eqn:gaussian_kernel}, a cubic phase kernel \eqref{eqn:cubic_kernel} and a squeezed kernel \eqref{eqn:squeezed_kernel}. We use 100 data samples each for the CVBM and the target distribution with 50 samples from shifted circuits to calculate MMD gradients. The shading around the plots indicates the standard deviation for each kernel after 5 runs of the training algorithm.}
    \label{fig:kernel_tests}
\end{figure}

Here, we explore the behaviour of the MMD estimator with two separate kernel mappings: the \textit{cubic phase kernel} $\phi_V$ \eqref{eqn:cubic_kernel} and the \textit{squeezed kernel} \eqref{eqn:squeezed_kernel} $\phi_S$. For data point $\xbs = x_1, x_2, ..., x_n$ we can implement $n$ modes, applying the given unitary with strength $x_i$ to the qumode indexed by $i$.
\begin{align}\label{eqn:cubic_kernel}
    \phi_V: x_i \rightarrow V(\gamma = x_i)\ket{0}_i := \ket{V_{x_i}}\\
    \phi_S: x_i \rightarrow S(\phi = x_i)\ket{0}_i := \ket{S_{x_i}} \label{eqn:squeezed_kernel}
\end{align}
The full feature map is given by the tensor product of each of these states, for example, with \eqref{eqn:cubic_kernel}, $\phi_V: \xbs \rightarrow \ket{V_{\xbs_i}} := \bigotimes_{i=1}^n\ket{V_{x_i}}$. We can then compute the overlap of two mapped states $\ket{V_{\xbs_i}}$ and $\ket{V_{\xbs_i'}}$ to extract the kernel as in \eqref{eqn:quantum_kernel}. Note that the overlap is computed by expressing the state in Fock space with a selected dimensional cut-off dependent on the number of qumodes required to estimate the kernel. \figref{fig:kernel_tests} demonstrates the behaviour of the MMD estimator during training for both of the quantum kernels as well as the classical Gaussian kernel given in \eqref{eqn:gaussian_kernel}.

While both quantum kernels exhibit a convergence to some minimum of the MMD loss, they do not show any improvement over the classical kernel with regards of the resulting distributions. The squeezed kernel, while more erratic, gives a better final result while the cubic kernel is very poor for this task (note the cut-off dimension used in simulating Fock space for both kernels was 15). This is by no means a reason to discount the use of quantum kernels in such algorithms, but does indicate that more analysis needs to be done for which types of data might benefit more from particular mappings.

\subsection{Noise} \label{ssec:noise}

\begin{figure}[ht]
     \centering
 \begin{subfigure}[t]{0.47\textwidth}
 \centering
     \begin{tikzpicture}
          \node (img)  {\includegraphics[width=\textwidth]{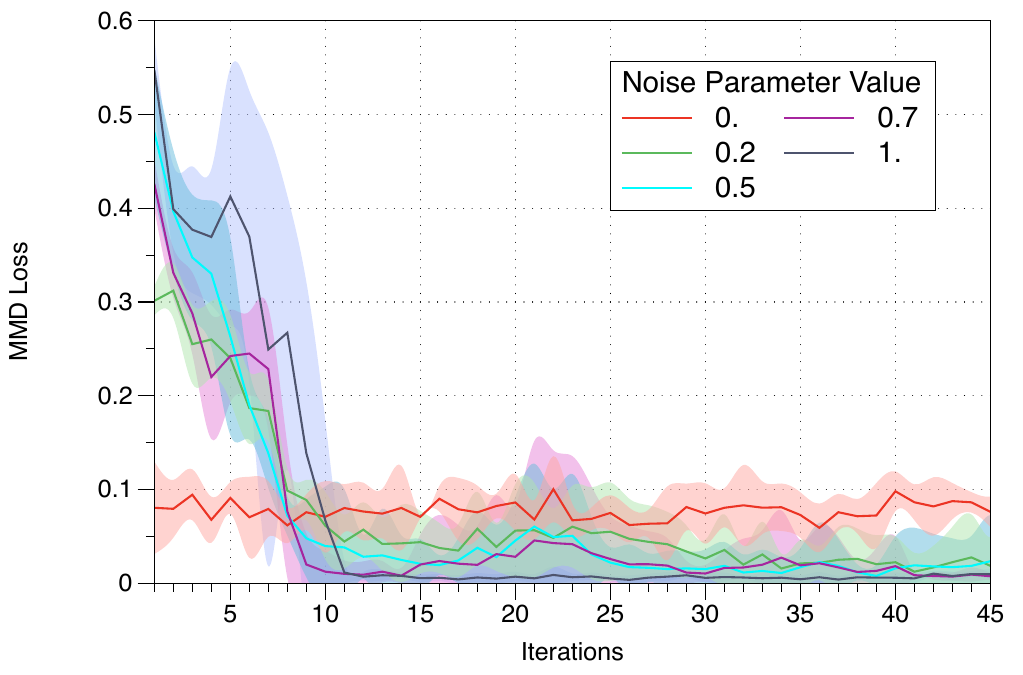}};
     \end{tikzpicture}
         \captionsetup{font=small, labelfont={bf}}
         \caption{Gaussian distribution parameterised by $S(\zeta)$ and $D(\alpha)$.}
         \label{subfig:gaussian_noise}
 \end{subfigure}
     ~ 
 \begin{subfigure}[t]{0.47\textwidth}
 \centering
     \begin{tikzpicture}
         \node (img)  {\includegraphics[width=\textwidth]{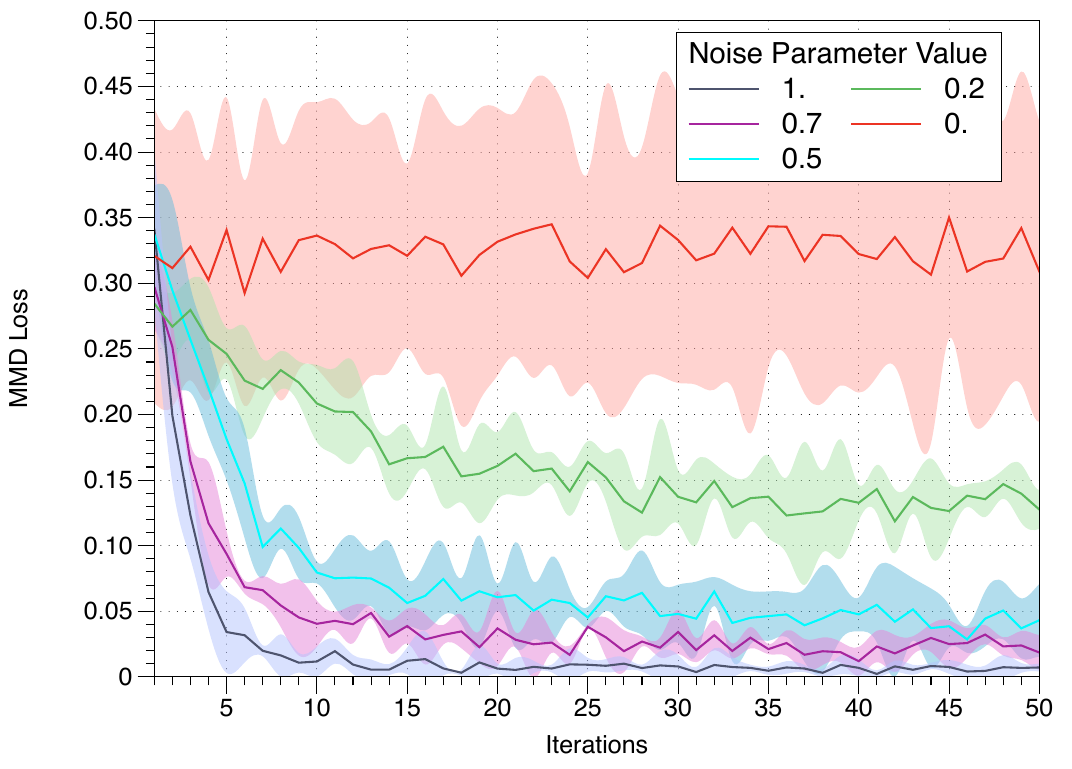}};
 \end{tikzpicture}
         \captionsetup{font=small, labelfont={bf}}
         \caption{Non-Gaussian distribution parameterised by $S(\zeta)$ and $V(\gamma)$.}
         \label{subfig:cubic_phase_gate_noise}
     \end{subfigure}
     \captionsetup{justification=justified,font=small, belowskip=-10pt, labelfont={bf}}
     \caption{Plots of MMD Loss for single-mode CVBMs learning a Gaussian (Figure~\ref{subfig:gaussian_noise}) and non-Gaussian (Figure~\ref{subfig:cubic_phase_gate_noise}) single-qumode state. The CVBM is coupled to a loss channel with different values of parameter $T$ (as given in \eqref{eqn:noise_channel}). We use 100 data samples each for the CVBM and the target distribution with 50 samples from shifted circuits to calculate MMD gradients as well as the classical Gaussian kernel \eqref{eqn:gaussian_kernel}. The shading around the plots indicates the standard deviation for 5 runs of the training algorithm.}\label{fig:noise_plots}
 \end{figure}

Finally, we investigate the effect of a simple noise model on the training of the CVBM. In general, noise in quantum systems can be modelled using a completely positive trace preserving (CPTP) map, $\mathcal{N}$, which can equivalently be expressed in an operator-sum~\cite{nielsen_quantum_2011} formalism, decomposed into Kraus operators. For the CVBM, we choose a simple noise model available in Strawberry Fields~\cite{killoran_strawberry_2018}, in order to study the effect of loss, whose Kraus representation, acting on a state, $\rho$, is modelled by:
\begin{equation}\label{eqn:noise_channel}
    \mathcal{N}^T(\rho) = \sum^{\infty}_{n = 0} E_n(T) \rho E_n(T)^{\dag}, 
\end{equation}
where
\begin{equation}\label{eqn:noise_channel2}
    E_n(T) = \bigg(\frac{1-T}{T}\bigg)^{\nicefrac{n}{2}} \frac{\hat{a}^n}{\sqrt{n!}} (\sqrt{T})^{\hat{a}^{\dag}\hat{a}} .
\end{equation}
This has the effect of coupling a mode $\hat{a}$ to another mode in the vacuum state $\hat{b}$ via the transformation:
\begin{equation}\label{eqn:noise_coupling}
    \hat{a} \xrightarrow{} \sqrt{T}\hat{a} + \sqrt{1-T}\hat{b}
\end{equation}
which is then traced out. The noise parameter $T$ represents energy transmissivity and $T = 1$ represents the identity map. For $T = 0$, the state is mapped to a vacuum state. \figref{fig:noise_plots} shows the effect of noise on the CVBM learning process for both a Gaussian and non-Gaussian quantum state.

As might be expected, the Gaussian state is significantly more robust to noise since the noise channel given by \eqref{eqn:noise_channel} couples the mode of interest to a vacuum state which is itself a Gaussian distribution. However, we can see that in the non-Gaussian case the algorithm adapts to some level of noise also, lending credence to using a CVBM in less-than-perfect experimental set-ups. Note that the target distribution data is assumed to have no noise channel affecting it as the assumption is that its source is unknown. However, the CVBM can also be used to determine the strength of noise on quantum data if its own coupling to a relevant noise channel can be manipulated.

\subsection*{Conclusion}

We have presented a new QML method within the CVQC model for learning continuous probability distributions. While there is still much to be explored (in particular, in the areas of selecting more apt kernels for the MMD estimator - be they classical or quantum in nature - as well as optimizing its implementation both on the classical and quantum hardware), it promises to be an interesting tool in exploring the relationship between the statistics of quantum experiments on CV states and the unitaries that `compile' them. It should be noted that the larger the number of samples from the target distribution, the more accurate the CVBM result is in producing relevant samples or in determining how close its circuit is to reproducing a particular state, particularly in the case of non-Gaussian states. It can also be implemented in learning classical distributions, particularly with high dimensionality, in cases where their parameterisation is difficult to capture with a classical model. As with many QML algorithms, the greatest advantage of this method is that it presents one with a complex model with an inherent large memory (Hilbert space) and a parameterisation that is mathematically complex yet easily accessible once run on a quantum device as well as being native to quantum physics.

\subsection*{Acknowledgements}\label{sec:acknowledgements}

\small{We thank Niraj Kumar and Ulysse Chabaud for useful comments. This work was supported by the Engineering and Physical Sciences Research Council (grant EP/L01503X/1), EPSRC Centre for Doctoral Training in Pervasive Parallelism at the University of Edinburgh, School of Informatics and Entrapping Machines, (grant FA9550-17-1-0055), the European Union's Horizon 2020 research and innovation program under grant agreement No. 817482 PASQuanS and the H2020-FETOPEN Grant PHOQUSING (GA no.: 899544).}

\bibliographystyle{unsrt}


\appendix

\section{Partial Derivatives for Non-Gaussian Transformations}\label{app_a}

\subsection{Unbiased Gradient Estimator of the Probability of a
CV Quantum Circuit}\label{app1}

Here we give a derivation for the gradient of the MMD cost function $\partial \mathcal{L}_{\textrm{MMD}}$ with respect to each of the parameters of the CV gates as given in Table \ref{table:gradients}. The derivation is largely based on work done in  \cite{liu_differentiable_2018} and gradients for Gaussian gates that were derived in \cite{schuld_evaluating_2019}. For the cubic phase gate $V(\gamma)$ and Kerr gate $K(\kappa)$, the partial derivative approximations originated as a part of this thesis and the derivation is given in Appendix \ref{app2}.

The form of the maximum mean discrepancy given in \cite{liu_differentiable_2018} is:

\begin{align}\label{eqn:probs}
    \mathcal{L}_{\textrm{MMD}} = \Big|\Big| \sum_{\boldsymbol{x}} p_{\theta}(\boldsymbol{x})\phi(\boldsymbol{x}) - \sum_{\boldsymbol{x}} \pi(\boldsymbol{x})\phi(\boldsymbol{x}) \Big|\Big|^2
\end{align}

where $p_{\theta}(\boldsymbol{x})$ and $\pi(\boldsymbol{x})$ are two probability distributions, while $\phi$ is a feature mapping, which can be a kernel function. When written in terms of expectation values of the samples, the above equations takes the exact same form as \eqref{eqn:mmd_cost_function} in the text.

If we then take a partial derivative of \eqref{eqn:probs} with respect to one of the parameters (say $\th_k$) that the distribution $p_{\theta}(\boldsymbol{x})$ depends on, we get:

\begin{align}\label{grad}
    \pdv{\mathcal{L}_{\mathrm{MMD}}}{\theta_k} = \sum_{\boldsymbol{x} ,\boldsymbol{y}} \kappa(\boldsymbol{x}, \boldsymbol{y})\Big(p_{\theta}(\boldsymbol{y})\pdv{p_{\theta}(\boldsymbol{x})}{\theta_k} + p_{\theta}(\boldsymbol{x})\pdv{p_{\theta}(\boldsymbol{y})}{\theta_k}\Big) - 2\sum_{\boldsymbol{x}, \boldsymbol{y}} \kappa(\boldsymbol{x}, \boldsymbol{y}) \pdv{p_{\theta}(\boldsymbol{x})}{\theta_k}\pi(\boldsymbol{y})
\end{align}

If we now treat $p_{\theta}(\boldsymbol{x})$ as a Born machine with observable $\boldsymbol{x}$, i.e.\@ the distribution we are interested in training, then we need to determine its derivative with respect to a particular parameter, $\flatfrac{\partial p_{\theta}}{\partial \theta_k}$. Luckily, we can refer to \cite{schuld_evaluating_2019} for the derivatives of each of the Gaussian gates with respect to their parameters as well as to Appendix \ref{app2} for the non-Gaussian ones. For the sake of demonstrating the method, we choose the displacement gate $D(\alpha)$. In \cite{schuld_evaluating_2019} its partial derivative is given as:
\begin{align}
    \partial_{\alpha}D(\alpha) = \frac{1}{2s}(D(\alpha + s) - (D(\alpha - s)), s \in \mathbb{R}
\end{align}

According to the same paper, this gradient corresponds to the gradient of an observable, which in our case would be the Homodyne measurement (Eq. \ref{eqn:homodyne_measurement}). Since the gradient of the Born machine is parametrised the same way we arrive at the partial derivative:
\begin{align}\label{partiala}
    \pdv{p_{\alpha}(\boldsymbol{x})}{\alpha} = \frac{1}{2s}\Big(p_{\alpha^+}(\boldsymbol{x}) - p_{\alpha^-}(\boldsymbol{x})\Big)
\end{align}

Where $\alpha^{\pm} = \alpha \pm s, s \in \mathbb{R}$. Substituting \eqref{partiala} into \eqref{grad} we get:
\begin{multline}
    \pdv{\mathcal{L}_{\textrm{MMD}}}{\alpha} = \frac{1}{2s}\Big(\sum_{\boldsymbol{x}, \boldsymbol{y}} \kappa(\boldsymbol{x}, \boldsymbol{y})p_{\alpha}(\boldsymbol{y})p_{\alpha^+}(\boldsymbol{x}) - \sum_{\boldsymbol{x}, \boldsymbol{y}} \kappa(\boldsymbol{x}, \boldsymbol{y})p_{\alpha}(\boldsymbol{y})p_{\alpha^-}(\boldsymbol{x}) + \\ \sum_{\boldsymbol{x}, \boldsymbol{y}} \kappa(\boldsymbol{x}, \boldsymbol{y})p_{\alpha}(\boldsymbol{x})p_{\alpha^+}(\boldsymbol{y}) - \sum_{\boldsymbol{x}, \boldsymbol{y}} \kappa(\boldsymbol{x}, \boldsymbol{y})p_{\alpha}(\boldsymbol{x})p_{\alpha^-}(\boldsymbol{y})\Big) - \\ \frac{1}{s}\Big(\sum_{\boldsymbol{x}, \boldsymbol{y}}\sum_{\boldsymbol{x}, \boldsymbol{y}} \kappa(\boldsymbol{x}, \boldsymbol{y})p_{\alpha^+}(\boldsymbol{x})\pi(\boldsymbol{y}) - \sum_{\boldsymbol{x}, \boldsymbol{y}} \kappa(\boldsymbol{x}, \boldsymbol{y})p_{\alpha^-}(\boldsymbol{x})\pi(\boldsymbol{y})\Big)
\end{multline}

Then we can use the symmetric condition of the kernel, $\kappa(\boldsymbol{x}, \boldsymbol{y}) = \kappa(\boldsymbol{y}, \boldsymbol{x})$ to arrive at the form of the gradient given in Table~\ref{table:gradients}. The same method can be applied to each of the parameters of each of the CV gates in order to derive the rest of the gradients given in Table~\ref{table:gradients}.

\subsection{Partial Derivatives for Non-Gaussian Transformations}\label{app2}

While in \cite{schuld_evaluating_2019} the gradients that are derived for Gaussian gates are analytic and based on their decomposition into covariance matrices in phase space, no such simplification was possible for the non-Gaussian gates:
\begin{align}
    V(\gamma) = \textnormal{exp}\Big(i\frac{\gamma}{3\hbar}\hat{x}^3\Big)
\end{align}
\begin{align}
    K(\kappa) = \textnormal{exp}(i\kappa \hat{n}^2)
\end{align}

Luckily, these unitaries exhibit certain properties which can be exploited in order to derive an approximate gradient. We take the cubic phase gate $V(\gamma)$ as an example.

First, we notice that regardless of what dimension we choose to truncate the unitary matrix that describes the gate at, the form of it will always look like:
\begin{align}\label{matrixform}
    V(\gamma) \approx \begin{pmatrix}
  e^{i\gamma x_{11}} & e^{i\gamma x_{12}} & \cdots \\
  e^{i\gamma x_{21}} & e^{i\gamma x_{22}} & \cdots \\
  \cdots & \cdots & \cdots
 \end{pmatrix}
\end{align}

where the terms $(x_{11}...x_{NN})$ come from the $N-$dimensional matrix of the operator $\hat{x}^3 = \Big(\sqrt{\frac{\hbar}{2}}(\hat{a} + \hat{a}^{\dagger})\Big)^3$ and whose values will vary slightly depending on the dimension at which we truncate the operators $\hat{a}$ and $\hat{a}^{\dagger}$. The derivative of the cubic phase gate with respect to its parameter $\gamma$ is then of the form:

\begin{align}
    \partial_{\gamma} V(\gamma) \approx i \begin{pmatrix}
  x_{11}e^{i\gamma x_{11}} & x_{12}e^{i\gamma x_{12}} & \cdots \\
  x_{21}e^{i\gamma x_{21}} & x_{22}e^{i\gamma x_{22}} & \cdots \\
  \cdots & \cdots & \cdots
 \end{pmatrix}
\end{align}

Now we can use the hyperbolic function identity:
\begin{align}
    \sinh(x) = \frac{1}{2}(e^x - e^{-x})
\end{align}

And the fact that for $x \ll 1$, $\sinh(x) \approx x$ to write out the partial derivative as a linear combination of the cubic phase gate with shifted parameters:
\begin{align}
    \partial_{\gamma} V(\gamma) \approx \frac{i}{2s}(V(\gamma + s) - V(\gamma - s)),\, s \ll 1
\end{align}

\noindent This is in a form that is well-suited to be implemented by the gradient of the MMD cost function as shown in Appendix \ref{app1}. The error $\sinh(x) \approx x$ scales as $O(x^3)$ and so for sufficiently small values of $x$ the gradient approaches the exact analytical solution.

When we turn to the Kerr gate $K(\kappa)$, we find that we can write it out in a similar form to the one in \eqref{matrixform} and thus we are left with a partial derivative given by:

\begin{align}
    \partial_{\kappa} K(\kappa) \approx \frac{i}{2s}(K(\kappa + s) - K(\kappa - s)),\, s \ll 1
\end{align}

\end{document}